\documentclass[twocolumn,
showpacs,
preprintnumbers,
amsmath,
amssymb,
amsfonts,
a4paper,
aps,
citeautoscript,
prb,
superscriptaddress
]{revtex4}

\usepackage{times}
\usepackage[OT1]{fontenc}
\usepackage{graphicx}
\usepackage{rotating}

\def\tlse{Laboratoire de Physique Th\'eorique, IRSAMC, Universit\'e Paul
Sabatier, CNRS UMR5152,\\ 118 Route de Narbonne, 31400 Toulouse, France.}

\def\lyon{Laboratoire de Physique de l'\'Ecole Normale Sup\'erieure de
Lyon, ENS-Lyon, CNRS UMR5672,\\ 46 All\'ee d'Italie, 69007 Lyon, France.}

\def\irvine{Department of Physics and Astronomy, University of
California, Irvine, California 92697, USA.}

\def\fig{Fig.}
\def\Ham{\mathcal{H}}
\def\nh{n_h}
\def\E{\mathcal{E}}

\def\P{\mathcal{P}}
\def\S{\mathbf{S}}
\def\H{\textrm{H}}
\def\S{\mathbf{S}}

\newcommand{\moy}[1]{\langle{#1}\rangle}
\def\ups{\uparrow}
\def\downs{\downarrow}

\begin{document}

\title{Diamagnetism of doped two-leg ladders and
probing the nature of their commensurate phases}

\author{G.\ Roux}
\email{roux@irsamc.ups-tlse.fr}
\affiliation{\tlse}
\author{E.\ Orignac}
\affiliation{\lyon}
\author{S.\ R.\ White}
\affiliation{\irvine}
\author{D.\ Poilblanc}
\affiliation{\tlse}

\date{\today}




\begin{abstract}
  We study the magnetic orbital effect of a doped two-leg ladder in
  the presence of a magnetic field component perpendicular to the
  ladder plane. Combining both low-energy approach (bosonization) and
  numerical simulations (density-matrix renormalization group) on the
  strong coupling limit (t-J model), a rich phase diagram is
  established as a function of hole doping and magnetic flux. Above a
  critical flux, the spin gap is destroyed and a Luttinger liquid
  phase is stabilized. Above a second critical flux, a reentrance of
  the spin gap at high magnetic flux is found. Interestingly, the
  phase transitions are associated with a change of sign of the
  orbital susceptibility. Focusing on the small magnetic field regime,
  the spin-gapped superconducting phase is robust but immediately
  acquires algebraic transverse (i.e.  along rungs) current
  correlations which are commensurate with the $4k_F$ density
  correlations.  In addition, we have computed the zero-field orbital
  susceptibility for a large range of doping and interactions ratio
  $J/t$ : we found strong anomalies at low $J/t$ only in the vicinity
  of the commensurate fillings corresponding to $\delta = 1/4$ and
  $1/2$. Furthermore, the behavior of the orbital susceptibility reveals
  that the nature of these insulating phases is different: while for
  $\delta = 1/4$ a $4k_F$ charge density wave is confirmed, the
  $\delta = 1/2$ phase is shown to be a bond order wave.
\end{abstract}
\maketitle

\section{Introduction}

Ladder systems have proven to be remarkably interesting systems, both
as simple models exhibiting behavior similar to 2D systems, and as
systems exhibiting competition between several types of ground states.
Theoretical models of doped ladders display a large superconducting
(SC) phase\cite{Dagotto1992, Hayward1995, Troyer1996, Balents1996,
Schulz1996, White2002} with $d$-wave pairing associated with the
presence of a spin gap, with a ground state which can be described
variationally as a short-ranged resonating valence bond
state\cite{Anderson1987}. Charge density wave (CDW) correlations
compete with the pairing correlations\cite{Nagaosa1995, Schulz1996,
White2002, Carr2002}. The phase diagram of the isotropic t-J model was
sketched in Ref.~~\onlinecite{White2002} and displays, in addition to
this competition, insulating phases for the particular
commensurate dopings $\delta = 1/4$ and $1/2$.  Another competition
exists between the superconducting phase and an orbital
antiferromagnetic flux (OAF) phase~\cite{Nersesyan1991,Schulz1996,
Scalapino2001, Tsutsui2001, Schollwock2003, Fjaerestad2006} which has
been addressed in different ladder models by studying transverse
current correlations which display a quasi-long range order in the OAF
phase.

\begin{figure}[t]
\centering
\includegraphics[width=0.75\columnwidth,clip]{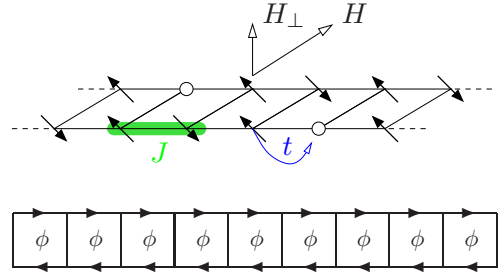}
\caption{(Color online) The isotropic t-J ladder under magnetic
field. If the magnetic field has a component perpendicular to the ladder plane, 
a flux $\phi$ passes through each plaquette. Below is
the gauge of the Peierls substitution with opposite phases $\pm
\phi / 2$ along the legs.}
\label{fig:schema}
\end{figure}

Ladders are also among the simplest systems through which a magnetic
flux can pass (see \fig~\ref{fig:schema}). When a magnetic field $\H$
is applied to an electronic system, it couples to both the spin of the
electron, via Zeeman effect, and to the charge degree of freedom, via
orbital effect. The total magnetic susceptibility of a real material
splits into various contributions\cite{Kjeldaas1957} $\chi =
\chi^{\text{spin}} + \chi^{\text{orb}}_{\text{cond}} +
\chi^{\text{orb}}_{\text{core}}$, where $\chi^{\text{spin}}$ is Pauli
susceptibility and $ \chi^{\text{orb}}_{\text{cond}}$ and
$\chi^{\text{orb}}_{\text{core}}$ are the orbital susceptibilities of
the conduction and the core electrons.
$\chi^{\text{orb}}_{\text{core}}$ must be evaluated from local atomic
orbital and we will neglect it in what follows.
$\chi^{\text{orb}}_{\text{cond}}$ is usually difficult to evaluate
because one has to precisely describe the evolution of the whole band
structure with magnetic field\cite{Hebborn1959, Kohn1959}. In the
following, we will investigate $\chi^{\text{orb}}_{\text{cond}}$
within a single-orbital model. When the magnetic field is applied
parallel to the plane of the ladder, the orbital effect is suppressed
and only the Zeeman effect remains. The latter case has been discussed
in details in this system and it was shown that a doping-controlled
magnetization plateau and a large Fulde-Ferrell-Larkin-Ovchinnikov
phase are obtained\cite{Cabra2002, Roux2006, Roux2007}. Since the
ladder possesses a spin gap, when the magnetic field is not in the
ladder plane, the orbital contribution may dominate the spin
contribution in the total susceptibility at low temperature.

Early numerical investigations of the t-J model with magnetic orbital
effect on ladders and 2D lattices revealed a strong effect of the
magnetic field on the magnetic and pairing
properties~\cite{Albuquerque2005}, but the results were limited to
small systems. A bosonization study of a related model of spinless
fermionic ladders suggested the possibility of fractional excitations
and of an OAF phase induced by the magnetic field~\cite{Narozhny2005,
Carr2006}. Carr and Tsvelik\cite{Carr2002} studied the orbital effect
of the magnetic field on the interladder coupling using an effective
model to describe a single ladder. Lastly, it has been
predicted\cite{Orignac2001} that bosonic ladders could have
commensurate vortex phases at commensurate fluxes which would
represent a one-dimensional analogue of the two-dimensional vortex
phase.

The purpose of the present paper is to consider the effect of a
nonzero flux on the magnetic susceptibility on a single two leg
ladder, and also to investigate the effect of stronger fluxes on the
zero temperature phase diagram of the ladder. To this end, we combine
the bosonization technique and density-matrix
renormalization group\cite{White1992, White1993, Schollwock2005}
(DMRG, see Appendix~\ref{sec:flux} for details) to compute the phase
diagram and physical properties of a spin-1/2 fermionic ladder with
orbital effect. The results are presented in two parts. The first part
is devoted to the analysis of the phase diagrams as a function of the
flux $\phi$ per plaquette in the weak- and strong-coupling limits. The
second part is focused on the physics of the spin-gapped phase of
doped ladder at small flux which is more relevant to realistic
magnetic fields. In particular, we discuss the stability of the
insulating phases at $\delta = 1/4$ and $1/2$ present in the phase
diagram of the t-J model at zero flux. In the conclusion, we briefly
give considerations on experiments which are connected to these
results. For sake of clarity, we have relegated some of the technical
details to appendices.

\subsection*{Including the flux}

The magnetic flux couples to the kinetic part of the Hamiltonian
through Peierls substitution \cite{Peierls1933, Kohn1959,
Kohn1964}. In what follows, different hopping amplitudes along the
chains ($t_{\parallel}$) and between the chains ($t_{\perp}$) are
considered, and the Hamiltonian is:
\begin{eqnarray}
\nonumber 
\Ham_t     &=& -t_{\parallel} \sum_{i,\sigma} 
\left[ e^{i\phi/2} c^{\dag}_{i+1,1,\sigma} c_{i,1,\sigma} + h.c.\right] \\
\nonumber && -t_{\parallel} \sum_{i,\sigma}
\left[ e^{-i\phi/2} c^{\dag}_{i+1,2,\sigma} c_{i,2,\sigma} + h.c.\right] \\
\label{eq:hamiltonian} && -t_{\perp} \sum_{i,\sigma}
\left[c^{\dag}_{i,2,\sigma} c_{i,1,\sigma} + h.c.\right]
\end{eqnarray}
where $c_{i,l,\sigma}$ is the electron creation operator at site $i$
on leg $l$ with spin $\sigma$. $\phi$ denotes the dimensionless flux
per plaquette
\begin{equation}
\label{eq:flux}
\phi = \frac{e}{\hbar}\oint_{\square} A(x) dx = \frac{e}{\hbar} \H_{\perp} a^2,
\end{equation}
with $A(x)$ the vector potential which depends on the gauge choice,
$a$ the lattice spacing. The magnetic field breaks time-reversal and
chain exchange symmetries which, as expected, will have notable
consequences on the current properties of the system.  Symmetry and
periodicity considerations allow us to limit the study to flux
$0\le \phi \le \pi$.  Exchanging chains amounts to reversing the
direction of the magnetic field.  More details on the gauge and flux
quantization on a finite systems can be found in
Appendix~\ref{sec:flux}.  The unit of $\phi$ is $2\pi \phi_0$ with
$\phi_0 = h/e = 4.1357\times 10^{-15}~\text{T m}^2$. For experimental
considerations, a flux $\phi = 0.01 \pi$ already corresponds to a very
high magnetic field of $\H \sim 800$ T for a typical value $a =
4$~\AA~ of the Cu--O--Cu bond in a cuprate. The gauge chosen in
Eq.~(\ref{eq:hamiltonian}) is represented on \fig~\ref{fig:schema}.

\section{Phase Diagrams}

\subsection{Weak-coupling limit}
\label{sec:weak-coupling}

We first introduce interactions between electrons using the Hubbard
ladder in magnetic flux. The Hamiltonian comprises the kinetic term
$\Ham_t$ incorporating the flux and the on-site repulsion $U > 0$:
\begin{equation}
\label{eq:hubbard}
\Ham  = \Ham_t + U\sum_{i,p} n_{i,p,\uparrow}   n_{i,p,\downarrow}\,.
\end{equation}
According to the usual strategy\cite{Schulz1996}, we will consider
first the limit $U = 0$ and study the non-interacting band
structure. Then, we will turn on $U \ll t_{\parallel},t_\perp$ so that
the band structure is not deformed and obtain the different phases in
this weak-coupling limit. Let us begin with the discussion of the band
structure.
 
\begin{figure}[b]
\centering
\includegraphics[width=0.8\columnwidth,clip]{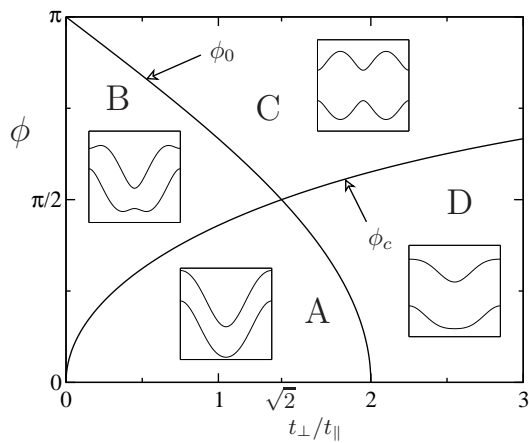}
\caption{The four typical shapes of the bands depending on the flux
and on the ratio $t_{\perp} / t_{\parallel}$. Critical fields $\phi_0$
(resp. $\phi_c$) signal the appearance of a double-well (band
gap). Note that the D phase always has only 2 Fermi points whatever
the filling and that the C phase \emph{when $\phi = \pi$} always has 4
Fermi points.}
\label{fig:Bands}
\end{figure}

The magnetic flux has a strong effect on the shape of the
bands. Indeed, it mixes the bonding ($0$) and antibonding ($\pi$)
bands which exist at zero flux. To emphasize the difference with the
zero field case, we call the two bands obtained at finite flux the
down ($d$) and up $(u)$ bands. Results on the band structure are
discussed in Refs.~~\onlinecite{Narozhny2005, Carr2006} and extended
to a finite and fixed filling in
Appendix~\ref{sec:non-interacting}. The band structure depends on the
flux $\phi$ and on the ratio $t_{\perp} / t_{\parallel}$ (see
\fig~\ref{fig:Bands}). We define two characteristic fluxes $\phi_c$
and $\phi_o$, both of them dependent on $t_{\perp} / t_{\parallel}$,
and such that above $\phi_c$, a double well appears in both bands and
above $\phi_0$, a band gap opens between the bands $u$ and $d$. Four
different possible shapes of the bands are obtained according to the
location of the flux with respect to $\phi_c$ and $\phi_o$.  The two
critical flux lines crosses at $(t_{\perp} / t_{\parallel} =
\sqrt{2},\phi = \pi / 2)$. By filling these bands, one can show (see
Appendix~\ref{sec:non-interacting}) that only situations with either
two or four Fermi points can occur.  The location of these Fermi
points and their respective Fermi velocities vary continuously with
the flux. In the rest of the paper, we will work only at fixed
electronic density (denoted by $n$) which will constrain the sum of
the Fermi momenta as a result of the Luttinger
theorem.\cite{Gagliardini1998}

\begin{figure}[t]
\centering
\includegraphics[width=0.9\columnwidth,clip]{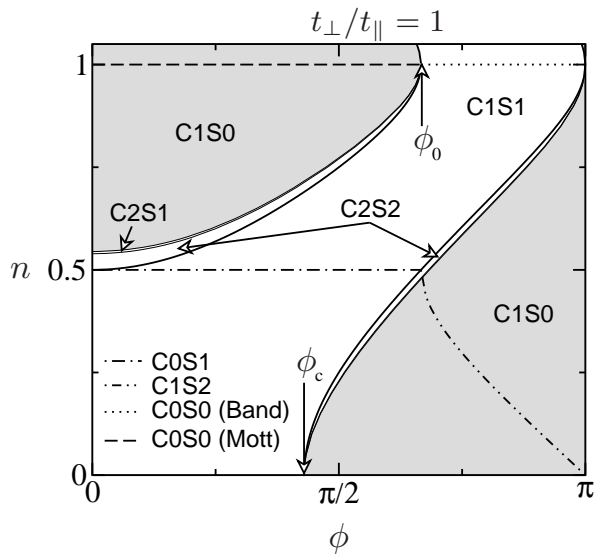}
\caption{Phase diagram in the weak-coupling limit for an isotropic
ladder restricted to fillings $0 \leq n \leq 1$. Phases with 4
(resp. 2) Fermi points fall generically into the C1S0 (resp. C1S1)
class. At half-filling, interactions will drive the system into a Mott
insulating phase for $\phi < \phi_0$ while a band insulating phase
occurs when $\phi > \phi_c$. Other phases can be found if the ratio of
the Fermi velocity is large (C2S1 and C2S2) and if the $d$-band Fermi
wave-vector is $\pi/2$ (C0S1 and C1S2).}
\label{fig:weak-coupling}
\end{figure}

Having obtained the non-interacting band structure, we add
interactions small enough not to perturb the band structure, following
the strategy of Refs.~~\onlinecite{Varma1985, Penc1990, Fabrizio1993,
Balents1996, Lin1997}. Adopting the usual notation C$p$S$q$ for a
phase with $p$ gapless charge modes and $q$ gapless spin modes, a
system with two Fermi points and repulsive interactions is expected to
be generically in a C1S1 phase, i.e. a Luttinger liquid state. With
four Fermi points and repulsive interactions, the system is
generically in a C1S0 phase, i.e. a Luther-Emery liquid which is the
universality class of usual doped two-leg ladders. The critical fields
at which the system changes from $4 \rightarrow 2$ or $2 \rightarrow
4$ Fermi points can by computed analytically (see
Appendix~\ref{sec:non-interacting}). Note that from
Refs.~~\onlinecite{Fabrizio1993, Balents1996, Lin1997}, in the case of
4 Fermi points, other phases such as C2S1 or C2S2 appear when the
difference between the two Fermi velocities becomes sufficiently large
to prevent runaway of some coupling constants in the RG flow.  The
large velocity difference implies that these phases are in the
vicinity of the transition region between the C1S0 and the C1S1 phase,
where the Fermi velocity of the band that is emptying is going to
zero. Moreover, this also implies that both the C2S1 and C2S2 phases
have a very small extent near the transition
region.\cite{Fabrizio1993, Balents1996, Lin1997} The above
considerations apply for a system at a generic incommensurate
filling. At commensurate filling, umklapp interactions can be relevant
and lower the number of gapless modes.\cite{Fabrizio1993, Balents1996,
Lin1997} More specifically, at half-filling, an insulating phase C0S0
of the Mott type (resp. Band type) is expected for $\phi < \phi_0$
(resp. $\phi > \phi_0$). It is also possible to have an umklapp
interaction inside the bonding band (if its Fermi wavevector equals
$\pi/2$) leading to either a C0S1 or C1S2
phase~\cite{Balents1996}. The phase diagram of an isotropic ladder
resulting from these considerations is given on
\fig~\ref{fig:weak-coupling} where the main feature is a reentrance of
the C1S0 phase at high flux. The high flux C1S0 phase has a band
structure very similar to the one of the Hubbard chain with a
next-nearest hopping term\cite{Fabrizio1996, Daul1998} $t'$ for $t'
\gtrsim t/2$ and the same competing orders as the low flux C1S0 phase
as we will see. This phase diagram is generic for $t_{\perp} /
t_{\parallel} < 2$ except that $\phi_0 < \phi_c$ when $t_{\perp} /
t_{\parallel} > \sqrt{2}$. For $t_{\perp} / t_{\parallel} > 2$, the
C1S0 phase at low flux around half-filling disappears.

\subsection{Strong-coupling limit: numerical results on the t-J model}

We now let the interactions go to the strong-coupling regime $U \gg t$
where the Hubbard model~(\ref{eq:hubbard}) reduces to the t-J model
with $J \simeq 4t^2/U$. In this limit, we only use isotropic couplings
$t = t_{\parallel} = t_{\perp}$ and $J = J_{\parallel} = J_{\perp}$ so
that the t-J Hamiltonian simply reads
\begin{equation}
\label{eq:tJ-hamiltonian}
\Ham_{\text{t-J}} = \P \Ham_t \P
+\,J\sum_{\moy{i,j}}[\S_{i}\cdot\S_{j} - \frac{1}{4} n_{i} n_{j}]\,,
\end{equation}
where $\S_i$ is the spin operator and $n_i = c_{i,\sigma}^{\dag}
c_{i,\sigma}$ is the electronic density operator (leg index is omitted
in $\S_i$ and $n_i$). $\P$ is the Gutzwiller projector which prevents
double occupancy on a site. Observables are computed with DMRG for the
range of doping $0 < \delta = 1-n < 0.5$. The phase diagram will be
discussed for the special case $J/t = 0.5$ for which the system has
dominant superconducting fluctuations\cite{Hayward1995, White2002}.

\subsubsection{Orbital susceptibility}

\begin{figure}[t]
\centering
\includegraphics[width=0.75\columnwidth,clip]{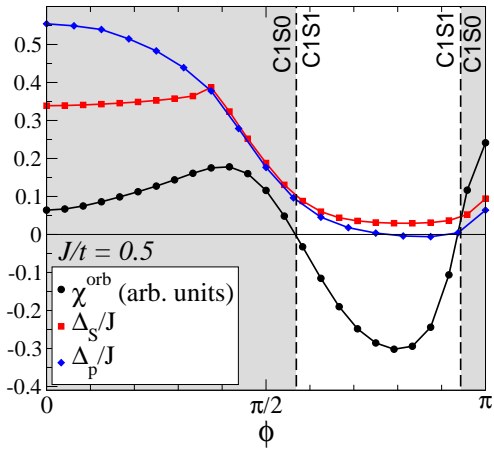}
\caption{(Color online) Orbital susceptibility, charge and spin gaps
for $J/t = 0.5$ and on the $\delta = 0.063$ line of
\fig~\ref{fig:phase-diagram}. The zeros of the susceptibility
precisely probe the different two phase transitions.}
\label{fig:comparison}
\end{figure}

\begin{figure}[t]
\centering
\includegraphics[width=0.85\columnwidth,clip]{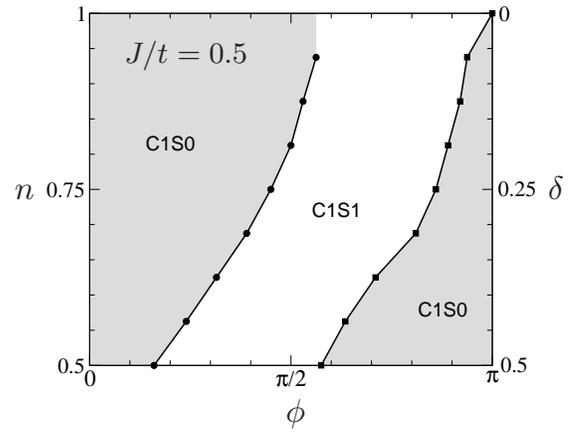}
\caption{Phase diagram of the t-J model for $J/t = 0.5$, for which the
system has dominant superconducting fluctuations at zero flux,
determined from the zeros of the susceptibility on a system with $L =
32$. Results are very similar to \fig~\ref{fig:weak-coupling} with a
transition to a C1S1 phase at intermediate flux and a reentrance to a
C1S0 phase at large flux.}
\label{fig:phase-diagram}
\end{figure}

The results for the non-interacting system of
Appendix~\ref{sec:non-interacting} show that the orbital
susceptibility plotted in \fig~\ref{fig:free-susceptibility} changes
sign at the transitions from $4 \leftrightarrow 2$ Fermi points with
sharp discontinuities (for $0 < \delta < 0.5$).  It is important to
note that the noninteracting orbital susceptibility contains
contributions from all the occupied states and not just those at the
Fermi level which control the low-energy properties. Therefore, such
connection of the change of sign of the orbital susceptibility and a
change in the number of Fermi point is not obvious.  Nevertheless, we
propose to extend this way of probing the phase diagram to the
interacting situation. Indeed, we can compute the screening current
$j_{\parallel}$ and its associated susceptibility $\chi^{\text{orb}}$
as a function of the flux $\phi$ using the definitions
\begin{equation}
\label{eq:screening-currents} 
j_{\parallel}(\phi) = -\frac 1 L \frac{\partial E_0}{\partial \phi}\quad
\text{and}\quad \chi^{\text{orb}}(\phi) = -\frac 1 L \frac{\partial^2 E_0}{\partial \phi^2}\,,
\end{equation}
in which $E_0(\phi)$ the ground-state energy and $L$ is the length of
the ladder. With this definition, $\chi^{\text{orb}}(\phi) > 0$
corresponds to orbital diamagnetism. The first relation is a
consequence of the Feynman-Hellman theorem and the second one results
from the definition of the susceptibility as $\partial
j_{\parallel}/\partial \phi$. This screening current can be related to
the mean value of the current operators $j_{1,2}$ along the two chains
by noting that $j_{\parallel}(\phi) = \moy{j_2-j_1}/2$. Numerically, these
quantities are computed directly from centered energy differences (to
minimize discretization effects)
\begin{eqnarray}
\label{eq:screening-currents-num}
j_{\parallel}(\phi) &=& -[E_0(\phi+d\phi) - E_0(\phi-d\phi) ] /(2Ld \phi)\\
\chi^{\text{orb}}(\phi) &=& [j_{\parallel}(\phi+d\phi) - j_{\parallel}(\phi-d\phi)] / (2d \phi)\,,
\end{eqnarray}
using the conditions $j_{\parallel}(0) = j_{\parallel}(\pi) = 0$ (see
Appendix~\ref{sec:flux}) and the right and left derivatives for
$\chi^{\text{orb}}(0)$ and $\chi^{\text{orb}}(\pi)$. These quantity
are easy to compute numerically and are found to have small finite
size effects for $J/t = 0.5$.  The effect of interactions is to smooth
the discontinuities at the transitions but we still expect that the
sign-changes of the susceptibility do correspond to transitions
between C1S0 and C1S1 phases even in the strong-coupling limit (see
\fig~\ref{fig:comparison}). The phase diagram obtained from this
ansatz is consistent with the behavior of other observables such as
spin-spin correlation functions as will be seen in the next
paragraphs. Thus, we can sketch on \fig~\ref{fig:phase-diagram} a
phase diagram similar to the one of \fig~\ref{fig:weak-coupling} for
the t-J model with $J/t = 0.5$. Compared with the weak-coupling phase
diagram, the C1S1 phase is slightly wider at low doping but thinner
for $\delta \simeq 0.5$.  Even in the presence of strong interactions,
the overall shape of the phase diagram is not affected (although the
precise location of the phase boundaries in the $\delta,\phi$ plane
does depend on $U/t$ or $J/t$).  Hence, for typical densities $0.5 < n
< 1$ in the isotropic t-J model, the leading effect which governs the
phase diagram is the change of the band structure under the applied
flux. Lastly, one must note that the C1S0 phase persists at small
flux at quarter-filling $\delta = n = 0.5$ in contrast to
Fig.~\ref{fig:weak-coupling}. This can be qualitatively explained by
noting that renormalization group studies on coupled chains have shown
that interactions reduce the interchain hopping integral $t_{\perp}$
with respect to its non-interacting value\cite{Bourbonnais1991,
Bourbonnais1986}.

\subsubsection{Elementary excitations : pairing energy and spin gap}

The elementary excitations of a doped two-leg ladder are either the
creation of a magnon, which cost is the spin gap $\Delta_S$, or the
creation of two quasi-particles by breaking a Cooper pair, which cost
defines the pairing energy $\Delta_p$. Following
Refs.~~\onlinecite{Roux2005, Roux2007}, we compute them numerically
from the definitions
\begin{eqnarray}
\label{eq:spin-gap}
\Delta_S &=& E_0(\nh,S^z=1) - E_0(\nh,S^z=0) \\
\nonumber
\Delta_p &=& 2E_0(\nh-1,S^z=1/2) \\
\label{eq:pairing-energy}
         & & - E_0(\nh,S^z=0) - E_0(\nh-2,S^z=0)
\end{eqnarray}
with $E_0(\nh,S^z)$ the ground state energy of a system with $\nh$
holes and spin $S^z$ along the $z$ axis. Since one can always have a
$S^z = 1$ state by breaking a Cooper pair, the condition $\Delta_S <
\Delta_p$ must be satisfied in the thermodynamic limit. For $\phi=0$
and $J/t = 0.5$, it is known\cite{Roux2005} that the pair-breaking
excitation is larger than the magnon excitation (lowest triplet
excitation). These elementary gaps as a function of the flux are
displayed on \fig~\ref{fig:comparison}. We observe a decrease of the
pairing energy with $\phi$ toward 0 at the critical flux corresponding
to the onset of the C1S1 phase. In the C1S1 phase, we have a metallic
Luttinger liquid phase with zero pairing energy. The cancellation of
the pairing energy is thus the result of a band emptying mechanism and
should not be confused with a magnetic superconducting critical field
$H_{c2}$ which would correspond to a high density of vortices in a 2D
(albeit anisotropic) superconducting material. Indeed, we do not
observe a $H_{c1}$ superconducting critical field or commensurate
vortex phases as in the model of bosonic of
Ref.~~\onlinecite{Orignac2001}. A situation corresponding to a true
$H_{c2}$ critical field might rather be a small flux through an array
of coupled ladder. In this respect, the approach of
Ref.~~\onlinecite{Carr2002} copmutes correctly the $H_{c2}$ field up
to a few approximations.

The spin gap increases at low magnetic field until it crosses
$\Delta_p$ (see \fig~\ref{fig:comparison}). From local hole densities
(data not shown), the domain of hole pairs slightly shrinks which can
be interpreted as a reinforcement of the spin-liquid background at low
magnetic field, in agreement with exact diagonalization results
previously discussed\cite{Albuquerque2005}. For larger flux but still
in the C1S0 phase, the spin gap $\Delta_S$ becomes identical to the
pairing energy $\Delta_p$ and both decrease towards zero as the flux
is increased. The energy difference $\Delta_p - \Delta_S$ can be
interpreted as the energy of a bound state of a magnon and a hole
pair. This magnetic resonant mode was discussed previously at zero
flux by varying interactions\cite{Poilblanc2000, Poilblanc2004,
Roux2005} and its origin was related with the opening of doping
controlled magnetization plateaus\cite{Roux2006, Roux2007}. Thus, the
effect of adding Zeeman coupling at low flux ($\phi \lesssim \pi / 3$
for $J/t = 0.5$) would give very similar results to those of
Refs.~~\onlinecite{Roux2006, Roux2007} since the bound state survives
to rather high magnetic flux. Finally, a small spin gap is recovered
at high flux (near $\phi = \pi$) in agreement with the weak-coupling
limit predictions.

\begin{figure}[t]
\centering
\includegraphics[width=0.95\columnwidth,clip]{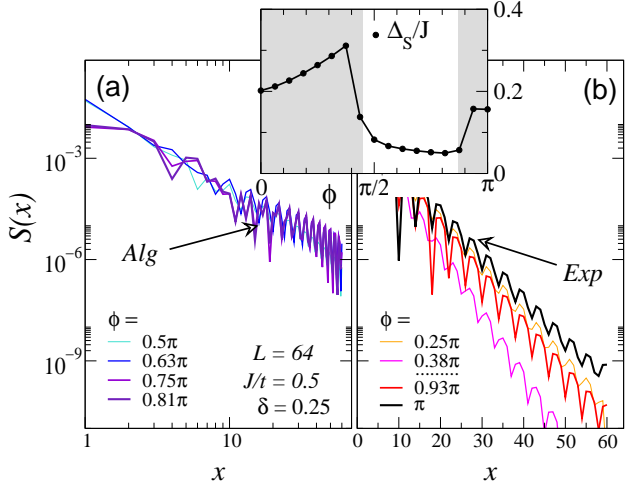}
\caption{(Color online) Spin correlations $S(x)$ for fluxes in the
C1S1 phase \textbf{(a)} and in the C1S0 phases at low and high fluxes
\textbf{(b)} in the phase diagram of \fig~\ref{fig:phase-diagram} on
the $\delta = 0.25$ line. \emph{Insert:} the behavior of the spin gap
computed on the same system using Eq.~(\ref{eq:spin-gap}). These
independent observables confirm the reentrance of the C1S0 phase.}
\label{fig:spin-correlations}
\end{figure}

To gain further insights on these excitations we have computed the
spin and pair correlation functions in the ground state. The spin
correlations $S(x) = \moy{\S(x) \cdot \S(0)}$ are short-range in a
spin-gapped phase with a correlation length $\xi \sim 1/\Delta_S$
which gives complementary estimation of the evolution of the spin gap,
particularly important when the pairing energy is smaller than the
spin gap. From \fig~\ref{fig:spin-correlations}, we find a similar
increase of the spin gap (smaller $\xi$) at small flux and algebraic
correlations in the C1S1 phase. The spin gap at high flux is again
recover with short-range spin correlations.

\begin{figure}[t]
\centering
\includegraphics[width=0.85\columnwidth,clip]{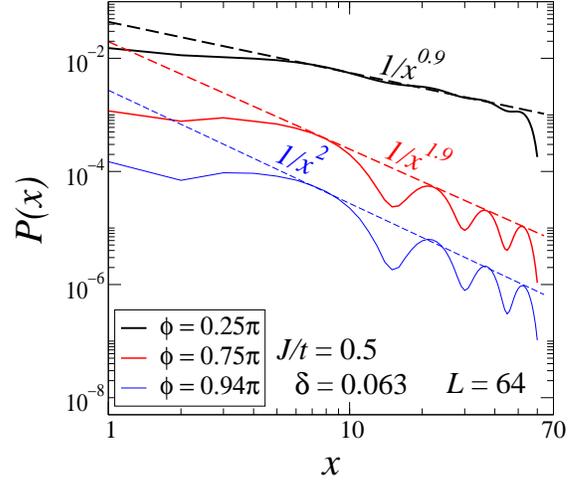}
\caption{(Color online) Superconducting correlations $P(x)$ for three
values of the flux corresponding to each phases encountered on the
$\delta = 0.063$ line of the phase diagram of
\fig~\ref{fig:phase-diagram}. The decrease exponent is much higher in
the C1S1 and high-flux C1S0 phases and the overall magnitude is
strongly reduced suggesting a metallic state. Note that
correlations display oscillations associated with a $4k_F$
contribution.}
\label{fig:pairing-correlations}
\end{figure}

The singlet operators on a rung defined by $\Delta(x) = c_{x,1,\ups}
c_{x,2,\downs} - c_{x,1,\downs} c_{x,2,\ups}$ gives the pairing
correlations $P(x) = \moy{\Delta(x) \Delta^{\dag}(0)}$. While $P(x)$
remains algebraic in the C1S0 and C1S1 phases, its overall magnitude
follows the pairing energy and is strongly reduced in the C1S1
phase. These correlations can be discussed from the bosonization
approach using the conventions and definitions of
Appendix~\ref{sec:bosonization}. We find that the low-energy dominant
term at wave-vector $q = 0$ in the low-field C1S0 phase reads
\begin{equation*}
\begin{split}
\Delta(x) \propto \sum_{\sigma} \sigma &\left[
          (a_d)^2 \psi_{d,R,\sigma}\psi_{d,L,-\sigma} 
         +(b_d)^2 \psi_{d,L,\sigma}\psi_{d,R,-\sigma} \right.\\
&\left.-(b_u)^2 \psi_{u,R,\sigma}\psi_{u,L,-\sigma} 
        -(a_u)^2 \psi_{u,L,\sigma}\psi_{u,R,-\sigma} 
\right].
\end{split}
\end{equation*}
For instance, the intraband terms read
\begin{equation}
\label{eq:bos-intraband-SC}
\psi_{p,R,\sigma}\psi_{p,L,-\sigma} \sim
e^{i [\theta_{c+} + p\theta_{c-} - \sigma (\phi_{s+} + p\phi_{s-})]}\,,
\end{equation}
with $p = \pm$ for $d/u$. From previous results\cite{Schulz1996}, we
know that in a C1S0 phase all the fields except $\phi_{c+}$ are
gapped, with $\langle \theta_{c-} \rangle =0$ and $\langle
\phi_{s+} \rangle = \pi / 2$, $\langle \phi_{s-} \rangle
=\pi/2$. These terms are thus algebraic with a decay exponent $1/(2K_{c+})$
which is the continuation of the zero-flux physics. We observe from
\fig~\ref{fig:pairing-correlations} that $K_{c+}$ increases with the
magnetic field but a precise evaluation is numerically difficult.

In the C1S1 phase, superconducting correlations are expected and found
to be algebraic with an exponent $K_c^{-1} + K_s$ but with a much
smaller amplitude, in agreement with a metallic phase.

The physical properties of the high-flux C1S0 phase are very similar
to that of the low-field C1S0 phase. Following the analysis by
Fabrizio\cite{Fabrizio1996} (see Appendix~\ref{sec:bosonization} for
notation), the pairing order parameter in this phase reads :
\begin{equation*}
\begin{split}
\Delta(x) \propto \sum_{\sigma} \sigma &\left[
          (a_1)^2 \psi_{1,R,\sigma}\psi_{1,L,-\sigma} 
         +(b_1)^2 \psi_{1,L,\sigma}\psi_{1,R,-\sigma} \right.\\
&\left.+(b_2)^2 \psi_{2,R,\sigma}\psi_{2,L,-\sigma} 
        +(a_2)^2 \psi_{2,L,\sigma}\psi_{2,R,-\sigma} 
\right].
\end{split}
\end{equation*}
which will give fluctuations with an exponent $1/(2K_{c+})$. Computing
the density and transverse current order parameters shows that they
have a $2K_{c+}$ decay exponent associated with the wave-vector
$2(k_{F,1} - k_{F,2}) = 2\pi n $. The competing orders are thus the
same as in the low field phase studied in
Sec.~\ref{sec:low-field}. The fact that the SC signal is small and
with a large exponent in \fig~\ref{fig:pairing-correlations} suggests
that the CDW fluctuations dominates in this strong-coupling regime
making the high flux C1S0 phase a spin-gapped metallic phase with
strong transverse current fluctuations.

\section{Low-field properties of the Luther-Emery phase}
\label{sec:low-field}

This section discusses the properties of the C1S0 Luther-Emery phase
at very low fluxes relevant to the experimental accessible magnetic
fields.

\begin{sidewaysfigure}[p]
\caption{(Color online) Local hole density and local currents in the
ground state of the C1S0 phase (at low flux). The area of the circles
is proportional to $h(x)-\delta$ with $h(x)$ the local density of
holes. An excess of hole is represented by empty circles while a lack
of holes is represented by full circles. The line thickness is
proportional to the current strength and the arrow gives the
direction. Note that the transverse current has been rescaled (by a
factor $(\times 5)$), the main screening currents are on the
chains. Interestingly, the orbits have a length scale $\delta^{-1}$
determined by the average hole density.}
\label{fig:locals}
\includegraphics[width=\textheight,scale=1.0]{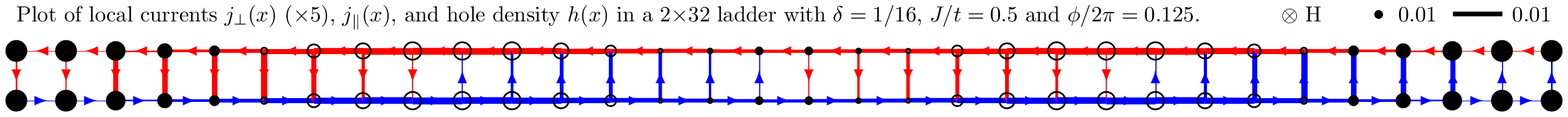}

\vspace{10mm}
\caption{(Color online) Local expectation values with the conventions
of Fig.~\ref{fig:locals} on the $n = \delta = 1/2$ line
(quarter-filling) in the bond-density-wave phase at low flux.}
\includegraphics[width=\textheight,scale=1.0]{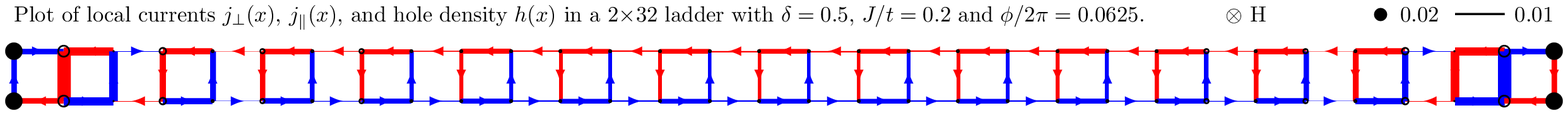}
\label{fig:All-locals-d0.5-J0.2}

\end{sidewaysfigure}

\subsection{Current densities and correlations}

With open boundary conditions used in DMRG, we have access to the
local density of holes and currents. The local hole density reads
$h(x) = 1 - \moy{ n(x)}$ while the mean values of local parallel and
transverse current operators are computed using the definitions
\begin{eqnarray}
\label{eq:local-currents}
\nonumber
j_{\parallel}^1(x) &=& it_{\parallel} [e^{i\phi/2}
c_{x+1,1,\sigma}^{\dag} c_{x,1,\sigma} - e^{-i\phi/2} c_{x,1,\sigma}^{\dag}
c_{x+1,1,\sigma} ]\\
\nonumber
j_{\parallel}^2(x) &=& it_{\parallel} [ e^{-i\phi/2}
c_{x+1,2,\sigma}^{\dag} c_{x,2,\sigma} - e^{i\phi/2} c_{x,2,\sigma}^{\dag}
c_{x+1,2,\sigma} ]\\
\label{eq:transverse-current}
j_{\perp}(x) &=& i t_{\perp} [c_{x,1,\sigma}^{\dag} c_{x,2,\sigma} -
c_{x,2,\sigma}^{\dag} c_{x,1,\sigma}]\,.
\end{eqnarray}
They are related to the current operator $j_p$ along the chain $p$
via $j_p = \frac 1 L \sum_x j_{\parallel}^p(x)$ and thus to the
screening current $j_{\parallel}(\phi)$. We have checked that Kirchhoff's
conservation law for charge currents is satisfied at each vertex of
the lattice. At zero flux, no currents are present in the ladder. When
the magnetic field is applied, time-reversal symmetry is broken and
local currents have a non-zero expectation value depicted in
\fig~\ref{fig:locals} for a $2\times 32$ ladder with 4 holes.  First,
the two hole pairs manifest themselves by two domains (areas with open
circles). The local screening currents develop inside these domains
and not at the edges of the ladder. Clearly, in the strong-coupling
limit where double occupancy is prohibited, the only domains in which
electrons can take advantage of the flux are the room left by holes.
This results in a periodic pattern for hole currents whose length
scale is exactly $\delta^{-1}$ ($= 16$ in \fig~\ref{fig:locals}). Such
a length scale is different from the usual magnetic length $\sqrt{
  \hbar / e H}$ governing orbits of Landau levels. A similar length
scale has been found in the study of OAF phases\cite{Fjaerestad2006}
while the current pattern is different from the one observed under
magnetic field.  Note that these diamagnetic currents are not related
to the Meissner effect expected in a superconductor or in a bosonic
ladder \cite{Orignac2001}. Actually, hole pairs are delocalized on the
two chains so that this pattern does not correspond to currents
\emph{of} pairs but rather to currents \emph{inside} pairs.

To study the nature of the current fluctuations, we have computed the
transverse current correlations
\begin{equation}
J(x) = \moy{j_{\perp}(x) j_{\perp}(0)} - \moy{j_{\perp}(x)}\moy{j_{\perp}(0)}\,,
\end{equation}
where we have subtracted the finite local expectations (as one would
do with density correlations). The main result of
Ref.~~\onlinecite{Scalapino2001} was the absence of algebraic
transverse current correlations in a C1S0 phase because of the strong
spin fluctuations associated with the spin gap. Here, although the
spin-gapped C1S0 phase survives at low flux, the situation is quite
different because the chain exchange symmetry is explicitly broken by
the magnetic field. Indeed, Eq.~(\ref{eq:transverse-current}) can be
rewritten in the basis of the $d,u$ bands, which leads in the
continuum limit to terms with different wave-vectors:
\begin{widetext}
\begin{eqnarray}
\label{eq:current-corr} 
j_{\perp}(x) & = & i t_\perp \left[(a_u^2-b_u^2) [ e^{-2i k_u x}
  \psi^\dagger_{u,R,\sigma} \psi_{u,L,\sigma} -  e^{2i k_u x}
  \psi^\dagger_{u,u,\sigma} \psi_{u,R,\sigma}] +  (a_d^2-b_d^2) [ e^{-2i k_d x}
  \psi^\dagger_{d,R,\sigma} \psi_{d,L,\sigma} -  e^{2i k_d x}
  \psi^\dagger_{d,L,\sigma} \psi_{d,R,\sigma}] \right. \nonumber \\ &&
\left.  + (b_u a_d + a_u b_d) [
  e^{-i (k_d+k_u) x} (\psi^\dagger_{u,R,\sigma} \psi_{d,L,\sigma}
  - \psi^\dagger_{d,R,\sigma} \psi_{u,L,\sigma} ) +
  e^{i (k_d+k_u) x} (\psi^\dagger_{u,L,\sigma} \psi_{d,R,\sigma} 
   - \psi^\dagger_{L,d,\sigma} \psi_{u,R,\sigma}) ] \right. \nonumber
 \\ && 
  + \left. (b_u b_d + a_u a_d)  [
  e^{-i (k_u-k_d) x} (\psi^\dagger_{u,R,\sigma} \psi_{d,R,\sigma}  - 
   \psi^\dagger_{d,L,\sigma} \psi_{u,L,\sigma})   
  + e^{i (k_u-k_d) x} (\psi^\dagger_{u,L,\sigma} \psi_{d,L,\sigma} -
   \psi^\dagger_{d,R,\sigma} \psi_{u,R,\sigma}) ] 
\right] 
\end{eqnarray}
\end{widetext}
where the coefficients $a_{d/u},b_{d/u}$ are defined in
Appendix~\ref{sec:bosonization}. We now turn to the bosonization
representation of the operators appearing in
Eq.~(\ref{eq:current-corr}). Terms with the lowest wave-vector $ k_d -
k_u$ contain operators of the form
\begin{equation}
\label{eq:bos-J0}
\psi_{u,R,\sigma}^{\dag} \psi_{d,R,\sigma} \sim  
e^{i[ -\phi_{c-}+\theta_{c-} - \sigma ( \phi_{s-} -\theta_{s-} ) ] }
\end{equation}
with $\sigma = (\ups,\downs) = \pm$. In the C1S0 phase, their
correlation functions decay exponentially as they involve the dual
fields $\theta_{s-}$ and $\phi_{c-}$ which are disordered.  For the
terms with the wave-vector $k_d + k_u$, we find similarly that
\begin{equation}
\label{eq:bos-J2}
\psi_{u,R,\sigma}^{\dag} \psi_{d,L,\sigma} \sim 
e^{i [ \phi_{c+} + \theta_{c-} + \sigma ( \phi_{s+} + \theta_{s-} ) ] }
\end{equation}
which also have short ranged correlation functions because of the
presence of the disordered field $\theta_{s-}$ in the bosonized
representation Eq.~(\ref{eq:bos-J2}).  This is exactly the same result
as in Ref.~\onlinecite{Scalapino2001} albeit extended to nonzero flux.
Without magnetic field, DMRG calculations\cite{Scalapino2001} showed
that the dominant wave-vector in the exponential signal was $k^{0} +
k^{\pi}$ rather than $k^{0} - k^{\pi}$ probably because the bosonized
expression of the corresponding Fourier component contains two
strongly fluctuating field for the latter term (see
Eq.~(\ref{eq:bos-J0}) , but only one (see Eq.~(\ref{eq:bos-J2})) for
the former term.  In the presence of a magnetic field, we see that the
terms in Eq.~(\ref{eq:current-corr}) corresponding to the wavevectors
$2k_u$ and $2k_d$ have a magnitude
\begin{equation*}
(b_p)^2 - (a_p)^2 \propto \phi
\end{equation*}
at small flux $\phi$, i.e. they exactly cancel for $\phi = 0$ but are
present once the magnetic field is turned on. These new term are
allowed by the symmetry reduction induced by the magnetic
field. Furthermore, they have the bosonized expression:
\begin{equation}
\label{eq:bos-J4}
\psi_{p,R,\sigma}^{\dag} \psi_{p,L,\sigma} \sim  
e^{i [ \phi_{c+} + p\phi_{c-} + \sigma ( \phi_{s+} + p\phi_{s-} ) ] }\,,
\end{equation}
in which the spin contribution has long range order, but the charge
contribution contains the dual field $\phi_{c-}$ leading to
exponential decay of the associated correlation function. Therefore,
all the ``$2k_F$'' contributions in the transverse current correlation
functions display an exponential decay.

\begin{figure}[b]
\centering 
\includegraphics[width=0.95\columnwidth,clip]{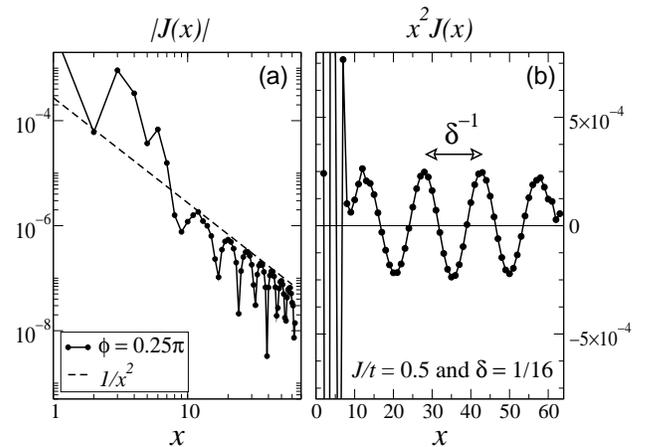}
\caption{\textbf{(a)} Transverse current correlations become algebraic
in the C1S0 phase once the magnetic field is turned on (same
parameters as in \fig~\ref{fig:locals}). \textbf{(b)} Demodulation of
the signal enables to extract clearly the wave-vector $2\pi \delta$ of
the correlations.}
\label{fig:current-correlations}
\end{figure}

In the expression (\ref{eq:current-corr}), the higher ``4k$_F$''
harmonics of the current are not taken into account. The reason for
this is that in the full Hilbert space, the $4k_F$ component of
$j_\perp$ is simply proportional to $c^{\dagger}_{3k_F}
c_{-k_F}$. However, in bosonization, a momentum cutoff is introduced,
and the high energy states which involve the creation or annihilation
of fermions with momentum farther from the Fermi momenta than the
cutoff are eliminated.  Thus, the expression of the current at lowest
order in the interaction $U/t$ in Eq.~(\ref{eq:current-corr}) cannot
contain any $4k_F$ contributions.  However, virtual processes in which
a fermion is created and annihilated far from the Fermi points give
contributions of higher order in $U/t$ to the $4k_F$ components that
only involve fermion operators close to the Fermi points.  Such
contributions can be derived in perturbation theory following the
approach in Refs.~\onlinecite{Penc1991,Tsuchiizu2001}.  In the case of
the transverse current, an interaction of the form $U c^{\dag}_{k_F}
c^{\dag}_{k_F} c_{-k_F} c_{3k_F}$ yields a perturbative contribution
proportional to $U/t\; c^{\dag}_{k_F} c^{\dag}_{k_F} c_{-k_F}
c_{-k_F}$ to the $4k_F$ component which involve only operators
belonging to the low energy subspace and thus cannot be
neglected. Such corrections can be viewed as a pair hopping or a
correlated hopping between the chains.  Thus, we expect to find a
$4k_F$ contribution to the transverse current of the form:
\begin{equation}
\label{eq:power-law-cur}
\psi_{d,R,\sigma}^{\dag} \psi_{d,L,\sigma}
\psi_{u,R,\sigma}^{\dag} \psi_{u,L,\sigma} \sim e^{i 2\phi_{c+}}\,,
\end{equation}
associated with the wave-vector $2(k^d + k^u) = 2\pi(1-\delta)$. The
$\moy{ j_{\perp,4k_F}(x) j_{\perp,4k_F}(0) }$ correlations have a
power-law decay with exponent $2K_{c+}$. This result is very similar
to the CDW fluctuations associated with the correlations $\moy{n(x)
n(0)}$ which, while being short-range at $2k_F$, also possess a $4k_F$
power-law decay\cite{White2002}. Note that these CDW correlations
contain terms analogous to (\ref{eq:bos-J0}, \ref{eq:bos-J2},
\ref{eq:bos-J4}) but with different prefactors. This is in good
agreement with numerical results for which we found algebraic
correlations with wave-vector $2\pi \delta$. The wave-length of the
correlations is again associated to the length scale $\delta^{-1}$ of
the local hole and transverse current patterns. We found numerically a
larger Luttinger exponent from the current correlations $K_{c+} \sim
1$ while superconducting correlations rather give $K_{c+} \sim 0.6$
for the same parameters. This difference, also found for charge
correlations\cite{Hayward1995}, could be attributed to the need for
virtual high energy processes to create the $4k_F$ correlations,
leading to somehow low and noisy signals. The behavior of the
bosonized operators is summarized in table~\ref{tab:summary}.

\begin{table}[t]
\centering
\caption{Summary of the bosonization result for operators in the low
field C1S0 phase (see Appendix~\ref{sec:bosonization} for
notations). We have $2k_F = \pi n$. If not short-range ("exp."
notation), we give the decay exponent of the associated
correlations. Numerically, $\moy{n(x) n(0)}$ and $\moy{j_{\perp}(x)
j_{\perp}(0)}$ are algebraic because they pick up the 4$k_F$ terms.}
\begin{tabular}{|c|c|c|}
\hline\hline
& \multicolumn{2}{c|}{ in the C1S0 phase } \\
\cline{2-3}
Operator & exponent & wave-vector \\
\hline\hline
$S^z(x)$       &  exp.  & $2k_F$ \\
\hline
$\Delta(x)$    & $1 /(2K_{c+})$ & 0\\
\hline
$n_{2k_F}(x)$         &  exp. & $2k_F$ \\
\hline
$n_{4k_F}(x)$       &  $2K_{c+}$ & $4k_F$ \\
\hline
$j_{\perp,2k_F}(x)$ & exp. & $2k_F$ \\
\hline
$j_{\perp,4k_F}(x)$ & $2K_{c+}$ & $4k_F$ \\
\hline\hline
\end{tabular}
\label{tab:summary}
\end{table}

\subsection{Zero-field susceptibility and the commensurate phases}

\begin{figure}[t]
\centering
\includegraphics[width=0.95\columnwidth,clip]{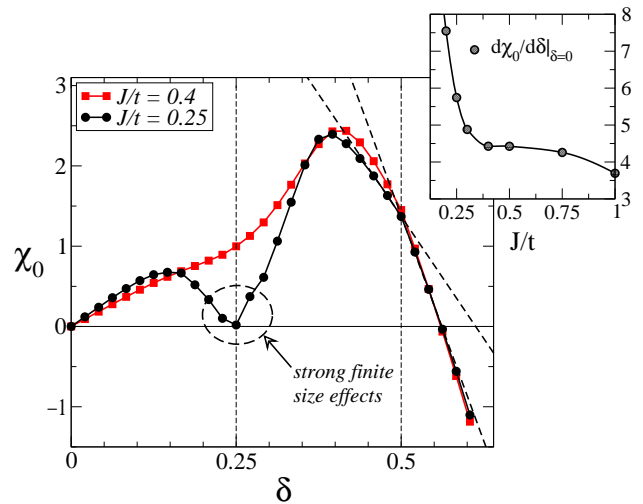}
\caption{(Color online) Zero field susceptibility $\chi_0$ as a
  function of doping $\delta$ and interactions parameters $J/t$ for a
  ladder with $L=48$. A strong reduction at low $J/t$ and
  commensurabilities $\delta = 1/4$ and $1/2$ are clearly visible. The
  computed susceptibility is nearly zero at $\delta = 1/4$ but subject
  ot finite size effects (see Appendix~\ref{sec:flux}). On the
  contrary, the $\delta = 1/2$ phase has a finite susceptibility.
  \emph{Insert:} the derivative $d\chi_0/d\delta\vert_{\delta = 0}$ as
  a function of $J/t$.}
\label{fig:zero-field-susceptibility}
\end{figure}

Since only very small flux per plaquette $\phi$ can be achieved
experimentally, we now focus on the $\phi \rightarrow 0$ limit of the
orbital susceptibility $\chi_0 \equiv \chi^{\text{orb}}(0)$ which is
calculated numerically from Eq.~(\ref{eq:screening-currents}). For the
non-interacting system, this quantity is finite and positive at
half-filling and then increases with doping (see
\fig~\ref{fig:free-susceptibility-n} of
Appendix~\ref{sec:non-interacting}). Once interactions are turned on,
this susceptibility is zero at half-filling because of the Mott
insulating state (see \fig~\ref{fig:zero-field-susceptibility} and
Appendix~\ref{sec:non-interacting} for a general discussion of the
susceptibility). When doped, the system acquires a susceptibility
roughly proportional to density of charge carriers with $\chi_0 \sim
\delta$. The proportionality coefficient decreases with $J/t$ which is
reminiscent of the fact that large $J$s reduce the mobility of holes
(see insert of \fig~\ref{fig:zero-field-susceptibility}). Compared
with the non-interacting result, the susceptibility is thus strongly
reduced by the interaction. When the ratio $J/t$ is lowered, a strong
reduction of $\chi_0$ is clearly visible for the hole commensurate
doping $\delta = 1/4$ up to finite size effects discussed in
Appendix~\ref{sec:flux}. This drop of the susceptibility increases
continuously as $J/t$ is lowered (data not shown, for larger $J/t$,
the convergence is better) so that we are confident that the
observation is not an artifact of the finite size effects. For $\delta
= 1/2$, a discontinuity of the slope is found but the susceptibility
remains finite in this phase (there, the finite size effects are
smaller). The occurrence of insulating CDWs was previously
studied\cite{White2002} in this part of the $(\delta,J/t)$ phase
diagram of the t-J model. However, only the $\delta = 1/4$ CDW phase
was discussed and \fig~\ref{fig:zero-field-susceptibility} suggests
that the $\delta = 1/2$ phase is of a different nature.

\begin{figure}[b]
\centering
\includegraphics[width=0.8\columnwidth,clip]{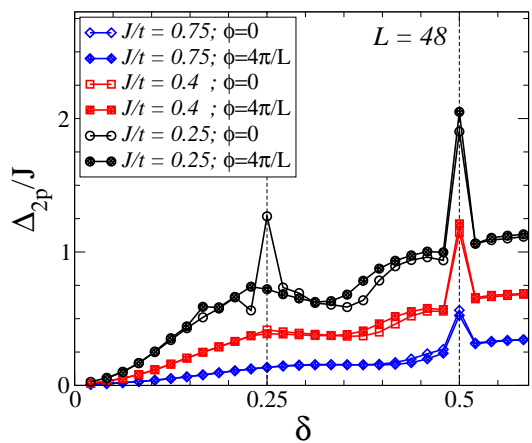}
\caption{(Color online) Two-particle charge gap as a function of
doping showing the emergence of insulating phases at
commensurabilities $\delta = 1/2$ and $\delta = 1/4$ at low $J/t$. The
smallest flux we have access to is enough to destroy the $\delta =
1/4$ CDW phase (see text for details), while the $\delta = 0.5$ BOW
phase is more stable. Data are computed with an extra $J = +0.3$ at
the two extremal rungs.}
\label{fig:charge-gap-2part}
\end{figure}

To study more precisely the occurrence of these phases, we have
computed the two-particle charge gap
\begin{eqnarray}
\label{eq:2-particle-charge-gap}
\Delta_{2p} =  E_0(\nh+2) + E_0(\nh-2) - 2E_0(\nh) 
\end{eqnarray}
as a function of doping and interaction at zero magnetic field and for
the lowest flux $4\pi/L$ we have access to. A system with pairs always
has a finite one-particle charge gap (or pairing energy), but is
insulating if the two-particle charge gap is also finite. The results
on a ladder of finite length $L=48$ are given on
\fig~\ref{fig:charge-gap-2part}. On the one hand, a strong
discontinuity at $\delta = 1/2$ is found even for rather large $J/t$
and the charge gap of this phase survives to nonzero flux (away from
the commensurability, $\Delta_{2p}$ is much smaller but finite because
of the finite length of the system). On the other hand, the $\delta =
1/4$ discontinuity is only found at small $J/t$ and is destroyed for
the lowest flux we can use. No other discontinuity of the two-particle
charge gap is found for the range of doping $0 \leq \delta \leq 0.6$.
The system with $\delta = 1/4$ has edge effects and we have added an
extra $J_{\perp} = +0.3$ on the two extremal rungs to control the
spinons at the edges as it was done in Ref.~\onlinecite{White2002}.
This phase is thus difficult to study under magnetic flux but it was
studied previously and it was proposed to be a four-fold degenerate
CDW phase with pairing and a small spin gap on the basis of the
behavior of the Friedel oscillations. These gaps were found to be
numerically very small\cite{White2002}. The observed sensitivity to
the flux is consistent with small gaps. Indeed, such a four-fold
degenerate phase is difficult to stabilize. Qualitatively, if pairs of
holes are well-formed on rungs, it is hard to generate an effective
long range repulsion between these pairs to stabilize a crystal of
hole pairs. On the contrary, if pairs are spread over a few rungs,
they can repel each other more easily but will have a smaller spin gap
and pairing energy. This latest picture of a pair of holes delocalized
on a plaquette every two plaquettes seems to be more suited to
describe this phase.

\begin{figure}[t]
\centering
\includegraphics[width=0.8\columnwidth,clip]{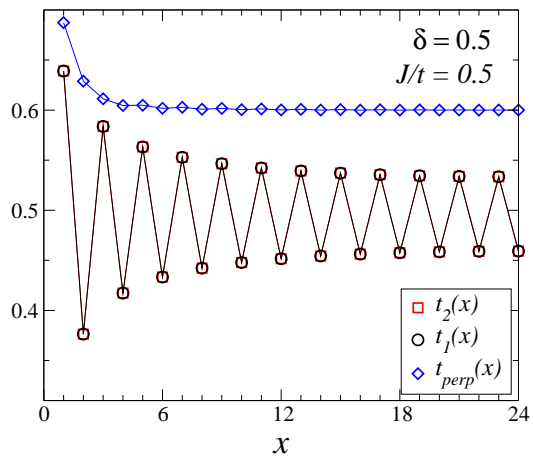}
\caption{(Color online) Local kinetic bonds $t_{\nu}(x)$ in the
$\delta = 1/2$ bond order wave (BOW) phase computed at zero magnetic
field. The parallel bond orders are strongly oscillating at
wave-vector $\pi$ while the transverse bond and the transverse kinetic
bond and electronic densities are uniform (see
Fig.~\ref{fig:All-locals-d0.5-J0.2}).}
\label{fig:BDW}
\end{figure}

In the insulating phase with $\delta = 1/2$ (quarter-filling), instead
of the pronounced Friedel oscillations obtained for $\delta = 1/4$, a
uniform electronic density (see \fig~\ref{fig:All-locals-d0.5-J0.2})
is found. However, if one computes the bond order parameters
$t_{\nu}(x)$ along the bonds at zero magnetic field by using the
definitions
\begin{eqnarray*}
\nonumber
t_{1,\parallel}(x) &=& t_{\parallel} \moy{
c_{x+1,1,\sigma}^{\dag} c_{x,1,\sigma} + c_{x,1,\sigma}^{\dag}
c_{x+1,1,\sigma}}\\
\nonumber
t_{2,\parallel}(x) &=& t_{\parallel} \moy{
c_{x+1,2,\sigma}^{\dag} c_{x,2,\sigma} + c_{x,2,\sigma}^{\dag}
c_{x+1,2,\sigma}}\\
t_{\perp}(x) &=& t_{\perp} \moy{c_{x,1,\sigma}^{\dag} c_{x,2,\sigma} +
c_{x,2,\sigma}^{\dag} c_{x,1,\sigma}}\,,
\end{eqnarray*}
one finds strong oscillations with a period of two lattice sites for
the $\parallel$ bonds making this phase an insulating bond order wave
(BOW) phase (see \fig~\ref{fig:BDW}). This is confirmed by the current
pattern under magnetic field found in
\fig~\ref{fig:All-locals-d0.5-J0.2} which has well-defined orbits
around plaquettes but small currents between plaquettes. The local
transverse current is staggered while the transerve bond density wave
order parameter is uniform. Such local orbits allow a finite orbital
susceptibility even if the system remains insulating because
$\moy{j_\perp(x) j_\perp(0)} \simeq \moy{j_\perp(1) j_\perp(0)} $ in
Eq.~(\ref{eq:general-orbital}) which is simply the local response on a
plaquette.

\section{Conclusion}

We have studied the effect of a magnetic flux through a doped two-leg
ladder by means of bosonization and DMRG calculations. As a function
of the flux, a rich phase diagram is observed with an intermediate
Luttinger liquid phase and the reentrance of the Luther-Emery phase at
high flux. Both in the weak- and in the strong-coupling limits, the
phase diagram is governed by the evolution of the band
structure. Focusing on the small field physics of the Luther-Emery
phase, we observe that local currents develop in the ladder inside the
hole pairs regions. Their typical length-scale $\delta^{-1}$ is
controlled by hole doping $\delta$. The transverse current
correlations also develop as soon as the magnetic field is turned on
with an algebraic behavior contrary to what was found without magnetic
field. Lastly, we have computed numerically the zero-field
susceptibility of the system as a function of the interaction
parameter $J/t$ and of hole doping. We found that insulating
commensurate phases at low $J/t$ exists only at dopings $\delta= 1/4$
and $1/2$ and that the two phases have different responses under
magnetic field. The contribution of the conduction electrons to the
orbital susceptibility might thus be useful to probe these phases
experimentally. Results on the $\delta = 1/4$ phase are consistent
with a four-fold degenerate ground state with a small pairing and
spin-gap, making it very sensitive to the flux. On the contrary, the
$\delta = 1/2$ phase appears to be a robust bond order wave phase with
a two-fold degenerate ground state. Despite its insulating nature,
this phase has a finite susceptibility due to local orbits of
electrons around plaquettes.
 
The ladder compound Sr$_{14-x}$Ca$_{x}$Cu$_{24}$O$_{41}$ (SCCO) was
the first non-square cuprate compounds showing superconductivity under
high pressure\cite{Uehara1996}. The presence of a spin gap in its
superconducting phase has been addressed experimentally
\cite{Mayaffre1998, Jerome2002, Fujiwara2003, Fujiwara2005} but no
consensus has risen on the actual nature of superconductivity in this
material. Another exciting feature of SCCO is the occurrence of charge
density waves at ambient pressure\cite{Osafune1999, Gorshunov2002,
Blumberg2002, Vuletic2003, Abbamonte2004, Gozar2005, Vuletic2005,
Choi2006, Rusydi2006, Rusydi2007}. Experiments\cite{Rusydi2006,
Rusydi2007} have suggested that CDW could appear at wave-vectors 1/3
and 1/5 which was discussed theoretically using a multiband charge
transfer model solved by Hartree-Fock approximation
\cite{Wohlfeld2007}. Here, we have showed that the t-J model on a
single ladder only displays 1/4 and 1/2 commensurabilities, as
proposed in Ref.~~\onlinecite{White2002}, and that orbital
susceptibility could help to understand the nature of these
commensurate phases.

We would like to point out that an interesting realization of
quasi-one dimensional systems in which magnetic flux can affect the
band structure is provided by carbon
nanotubes\cite{Ajiki1993}. Following the theoretical
prediction,\cite{Ajiki1993} experiments on multiwall nanotubes (where
notable fluxes can be achieved due to the large diameter of the
outmost shell) have shown that the band structure of these systems was
indeed sensitive to magnetic fluxes\cite{Zaric2004, Coskun2004,
Lassagne2007}. In the case of gapped zig-zag single wall nanotubes,
although the experimentally accessible fluxes through the tubes were
small, an effect on the conductance oscillations in the Fabry-Perot
regime could nevertheless be evidenced as a result of the lifting of
the degeneracy between two subbands. As there exists some evidence for
strong electronic correlations in carbon nanotubes\cite{Bockrath1999,
Egger2001, Ishii2003, Gao2004, Lee2004}, and as carbon nanotubes
possess some analogies with ladders\cite{Lin1998b}, an interesting
extension of the theoretical results developed in the present paper
would be to study quasi-one dimensional models mimicking more closely
carbon nanotubes. It would be particularly interesting to compute the
behavior of the Luttinger exponent controlling the Zero Bias Anomaly
as a function of the applied field.

\begin{acknowledgements}
GR would like to thank IDRIS (Orsay, France) and CALMIP (Toulouse,
France) for use of supercomputer facilities. GR and DP thank Agence
Nationale de la Recherche (France) for support. GR and SRW acknowledge
the support of the NSF under grant DMR-0605444.
\end{acknowledgements}

\appendix

\section{Formulas for the non-interacting system}
\label{sec:non-interacting}

Part of these results were first given in
Refs.~~\onlinecite{Narozhny2005, Carr2006}. We reproduce them for
clarity and notation conventions (which are different) and extend them
when necessary. In what follows $\alpha = t_{\perp} / t_{\parallel}$.

\subsection{Band structure}

At zero flux, the interchain coupling lifts the degeneracy between
chains giving birth to a bonding band $k_y = 0$ and anti-bonding band
$k_y = \pi$. The flux breaks the reflection symmetry between chains
and couples these $0$ and $\pi$ modes. We call down and up (with
labels $d/u$) the two bands in presence of a flux. It is
straightforward to diagonalize the non-interacting Hamiltonian by
taking the Fourier transform, which gives the energies
\begin{equation}
\label{eq:bands}
\E_{d/u}(k,\phi) =
-2t_{\parallel}\left\{ \cos k \cos \frac{\phi}{2} \pm \sqrt{\sin^2k
\sin^2\frac{\phi}{2} + \left(\frac{\alpha}{2}\right)^2} \right\}\,.
\end{equation}
The basis transformation can be written using coherence factors $a_k,
b_k > 0$
\begin{equation}\label{eq:basis-change}
  \begin{pmatrix}
    c_{k,1} \\
    c_{k,2} \\
  \end{pmatrix}=
  \begin{pmatrix}
    a_k & b_k \\
    b_k & -a_k \\
  \end{pmatrix}
  \begin{pmatrix}
    c_{k,d} \\
    c_{k,u} \\
  \end{pmatrix}\,,
\end{equation}
with
\begin{eqnarray}
\label{eq:coherences-factor-a}
a_k^2 &=& \frac 1 2 \left(1-\frac{\sin k \sin \frac \phi 2}
        {\sqrt{\sin^2 k \sin^2 \frac \phi 2 + \left(\frac \alpha 2\right)^2 }} \right)\\
\label{eq:coherences-factor-b}
b_k^2 &=& \frac 1 2 \left(1+\frac{\sin k \sin \frac \phi 2}
        {\sqrt{\sin^2 k \sin^2 \frac \phi 2 + \left(\frac \alpha 2\right)^2 }} \right)\,.
\end{eqnarray}
We have $a_{-k} = b_k$ and the factors depend on the wave-vector $k$
while at zero flux $a_k = b_k = 1 / \sqrt{2}$. For any finite flux and
$k>0$ we have $a_k < b_k$ with
\begin{equation}\label{eq:coherences-factor-diff}
b_k^2-a_k^2 = \frac{\sin k \sin \frac \phi 2}{\sqrt{\sin^2 k \sin^2
\frac \phi 2 + \left(\frac \alpha 2\right)^2 }}\,.
\end{equation}

\begin{figure}[b]
\centering
\includegraphics[width=5.8cm,clip]{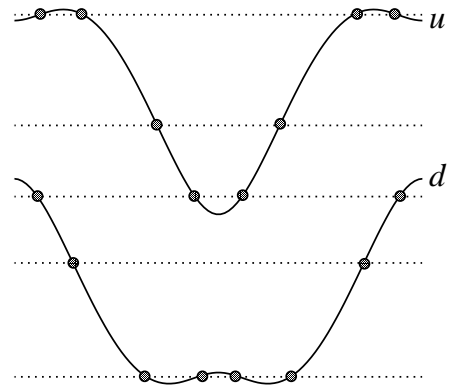}
\caption{Depending on the filling, the number of Fermi points can be
either 2 or 4 (sketched on the bands of the B phase of
\fig~\ref{fig:Bands}). Note that at low and high filling, the two
Fermi velocities have opposite signs while at intermediate filling,
they have the same sign.}
\label{fig:FermiPoints}
\end{figure}

Below are a few considerations on the band structure (\ref{eq:bands})
which are summarized in \fig~\ref{fig:Bands}:

(i) The condition to have a band gap is that $\max \E_d =
\E_d(\pi,\phi) < \min \E_u = \E_u(0,\phi)$ which gives $\frac \alpha 2
= \cos \frac \phi 2$. This condition can be reformulated as $\phi >
\phi_0$ with
\begin{equation}\label{eq:free-defining-phi0}
\sin \frac {\phi_0} 2 = \sqrt{1 - \left(\frac \alpha 2\right)^2 }\,.
\end{equation}

(ii) The condition to have a double well in the $\E_d$ band comes from
the sign of the second derivative at $k=0$ and is $\frac \alpha 2 \cos
\frac \phi 2 = \sin^2 \frac \phi 2$. This condition can be
reformulated as $\phi > \phi_c$ with
\begin{equation}\label{eq:free-defining-phic}
\sin \frac {\phi_c} 2 = \left[\frac{ \sqrt{\alpha^4 + 16 \alpha^2} -
\alpha^2 }{8}\right]^{1/2}\,.
\end{equation}
In this case, the wave-vectors corresponding to the minimum of the
down band and the maximum of the up band read
\begin{eqnarray}
k^d_{\text{min}}(\phi) &=& \arcsin \sqrt{ \sin^2 \frac{\phi}{2} -
\left(\frac{\alpha}{2}\right)^2 \cot^2 \frac{\phi}{2}}\\
k^u_{\text{max}}(\phi) &=& \pi - k^d_{\text{min}}(\phi)
\end{eqnarray}
which appears with a finite value.
 
(iii) The two curves intersection is $\alpha = \sqrt{2}$ for which we
have $\phi_{0/c} = \pi/2$. If $\alpha < \alpha_c$ we have $\phi_c <
\phi_0$, else $\phi_0 < \phi_c$.

(iv) The condition to empty the $u$ band is $\mu = \alpha - 2\cos
\frac \phi 2$ and is the same as the condition for which the $d$ band
is completely filled.

(v) We can show that only situations with two or four Fermi points
can occur: for monotonous bands ($\phi < \phi_c$), this is obvious. 
For non-monotonous dispersion ($\phi > \phi_c$), if $\phi >
\phi_0$, the two bands are not overlapping so we have either two or
four Fermi points. For any flux $\phi \in [\phi_c, \phi_0]$, we can
convince ourselves that since $\E^u(k,\phi) > \E^d(k,\phi)$
(for any $k$) and  since the up band has a unique maximum (at
$k^u_{\text{max}}(\phi)$) while the down band has a unique minimum (at
$k^d_{\text{min}}(\phi)$), it necessarily implies that only two or four
Fermi points are allowed (see \fig~\ref{fig:FermiPoints}).

\subsection{Filling the bands : finding Fermi points}

Fermi points $k_{F}^{d/u}$ are deduced from their relation to the
chemical potential $\mu$ from Eq.~(\ref{eq:bands}). If needed, this
equation can be inverted into
\begin{equation}
\label{eq:bands-cosk}
\begin{split}
\cos k_{F}^{d/u}(\mu,\phi) &= -\frac{\mu}{2t_{\parallel}} \cos
\frac{\phi}{2} \\
&\pm s_{d/u} \sqrt{ \left[ 1 -
\left(\frac{\mu}{2t_{\parallel}}\right)^2 \right] \sin^2
\frac{\phi}{2} + \left(\frac{\alpha}{2}\right)^2} \,.
\end{split}
\end{equation}
This latest is useful when working at fixed $\mu$. Note that,
depending on $\phi$ and $\mu$, the above equation can have two roots
labelled by $s_{d/u} = \pm 1$ for \emph{each} sector $d,u$ (see
\fig~\ref{fig:FermiPoints}). If there are two Fermi points in the same
band $p$, we use the notation $k_{F,1/2}^{p}$. The Fermi velocities
can be evaluated from
\begin{equation}\label{eq:free-energy-derivative}
v_F^{d/u}(k,\phi) = 2t_{\parallel} \sin k \left\{ \cos \frac{\phi}{2} \mp
\frac{\cos k \sin^2\frac{\phi}{2} }{\sqrt{\sin^2k \sin^2\frac{\phi}{2}
+ \left(\frac{\alpha}{2}\right)^2}} \right\}\,.
\end{equation}
Note that one of the two Fermi velocities is negative in case of
non-monotonous bands.

Working at fixed electronic density $n$, we have to relate the Fermi
points directly to $n$ using Eq.~(\ref{eq:bands}) and the Luttinger
sum rule (see \fig~\ref{fig:FermiPoints} for sketch of all possible
situations). If the system has only two Fermi points, then we have
either $k_{F}^d = \pi n$ or $k_{F}^u = \pi (n-1)$, where both
relations do not depend on the flux. When there are four Fermi points
and limiting the discussion to $n \leq 1$, we have either that:\\
\noindent (i) the bands are overlapping:
\begin{eqnarray}
\nonumber
k_F^u + k_F^d &=& \pi n \\
\nonumber
\sin\left(\frac{k_F^u - k_F^d}{2}\right) &=& \left[\frac{(\frac \alpha 2)^2 \cos^2 \frac \phi 2}
       {\sin^2(\frac{\pi n}{2}) - \sin^2 \frac \phi 2}+ \sin^2 \frac \phi 2 \right]^{1/2}\,.
\end{eqnarray}
\noindent (ii) there is a non-monotonous dispersion, for the down band:
\begin{eqnarray}
\nonumber
k_{F,1}^d - k_{F,2}^d &=& \pi n \\
\nonumber
\sin \left(\frac{k_{F,1}^d + k_{F,2}^d}{2}\right) &=&
\left[\frac{(\frac \alpha 2)^2 \cos^2 \frac \phi 2}
       {\sin^2(\frac{\pi n}{2}) - \sin^2 \frac \phi 2 }+ \sin^2 \frac \phi 2\right]^{1/2}\,.
\end{eqnarray}
These equations reproduce the correct result in the $\phi =0$ and $\alpha
= 0$ limits. It is also straightforward to compute the 4 critical
densities at which the number of Fermi points changes from 2 to 4 and
4 to 2 (see \fig~\ref{fig:weak-coupling}). By arranging them according
to $n_{c1} < n_{c2} < n_{c3} < n_{c4}$, we have
\begin{eqnarray}
\nonumber
 n_{c1} &=& \arccos(\cos \phi + \alpha \cos(\phi/2))/\pi \quad\text{if } \phi \in [\phi_c,\pi] \\
\nonumber
 n_{c2} &=& \arccos(\cos \phi - \alpha \cos(\phi/2))/\pi \quad\text{if } \phi \in [0,\phi_0] \\
\nonumber
 n_{c3} &=& 2 - n_{c2} \quad\text{if } \phi \in [0,\phi_0] \\
 n_{c4} &=& 2 - n_{c1} \quad\text{if } \phi \in [\phi_c,\pi] 
\end{eqnarray}
These equations can be inverted to give the critical flux $\phi^{2
\leftrightarrow 4}$ at which the transitions from 2 $\leftrightarrow$
4 Fermi points occur:
\begin{equation}
\label{eq:phi24}
\cos \frac {\phi^{2 \leftrightarrow 4}}{ 2} = \frac 1 2 \left [ \pm
\frac {\alpha} 2 + \sqrt{\left(\frac {\alpha} 2\right)^2 + 4
\cos^2\left(\frac{\pi n} 2\right)} \right ]
\end{equation}
this last expression is useful to check that the change of sign of the
susceptibility is associated with these transitions.

\subsection{Orbital susceptibility}

\begin{figure}[t]
\centering
\includegraphics[width=0.8\columnwidth,clip]{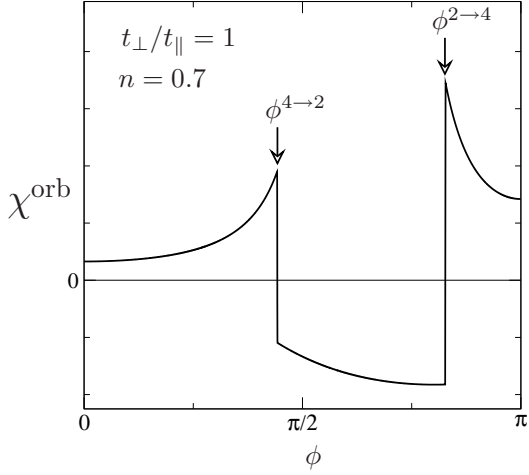}
\caption{Susceptibility as a probe to the number of Fermi points for
$0.5 < 1-\delta < 1.0$ and $t_{\perp} =
t_{\parallel}$. $\phi^{2\leftrightarrow 4}$ indicate the transitions
from 4 to 2 and 2 to 4 Fermi points.}
\label{fig:free-susceptibility}
\end{figure}

Knowing the location of the Fermi points , we can compute quantities
integrated over the bands such as the total energy $E_0(\phi)$, the
screening current $j_{\parallel}(\phi)$ and the associated orbital
susceptibility from Eqs.~(\ref{eq:screening-currents}). One
contribution to this current from electrons of momentum $k$ is
$j_{\parallel}^{d/u}(k,\phi) = \partial \E_{d/u}(k,\phi)/\partial
\phi$, which reads:
\begin{equation}
\label{eq:k-currents} 
j_{\parallel}^{d/u}(k,\phi) = -t_{\parallel}\left\{ \cos k \sin
\frac{\phi}{2} \mp \frac 1 2 \frac{\sin^2k \sin \phi }{\sqrt{\sin^2k
\sin^2\frac{\phi}{2} + \left(\frac{\alpha}{2}\right)^2 }} \right\}\,,
\end{equation}
and similarly, the corresponding contribution to the susceptibility reads
\begin{equation}
\label{eq:k-susceptibility} 
\begin{split}
\chi^{\text{orb}}_{d/u}(k,\phi) = -\frac{t_{\parallel}}{2}
&\left\{ \cos k \cos \frac{\phi}{2} 
\mp \frac{\sin^2 k \cos\phi }{\left[ \sin^2k \sin^2\frac{\phi}{2} +
\left(\frac{\alpha}{2}\right)^2 \right]^{1/2}} \right.\\
& \left. \pm \frac 1 4
\frac{\sin^4 k \sin^2 \phi }{\left[\sin^2k \sin^2\frac{\phi}{2} +
\left(\frac{\alpha}{2}\right)^2 \right]^{3/2}} 
\right\}\,.
\end{split}
\end{equation}
It is important to remark that additional contributions come from the
derivatives $\partial k^p(\phi) / \partial \phi$ and $\partial^2
k^p(\phi) / \partial \phi^2$ because, in the case of four Fermi
points, the location of the Fermi points depends on the flux. Thus,
Eqs.~(\ref{eq:screening-currents}) can be computed either analytically
or numerically. A typical plot of the integrated
$\chi^{\text{orb}}(\phi)$ is given on
\fig~\ref{fig:free-susceptibility} which shows that its
discontinuities are associated with the $2 \leftrightarrow 4$ Fermi
points transitions of Eqs.~(\ref{eq:phi24}). We have checked that the
latest result is valid only for $0.5 < n = 1 - \delta< 1.0$ when
$\alpha = 1$ (see zero-field susceptibility below).

\begin{figure}[t]
\centering
\includegraphics[width=0.8\columnwidth,clip]{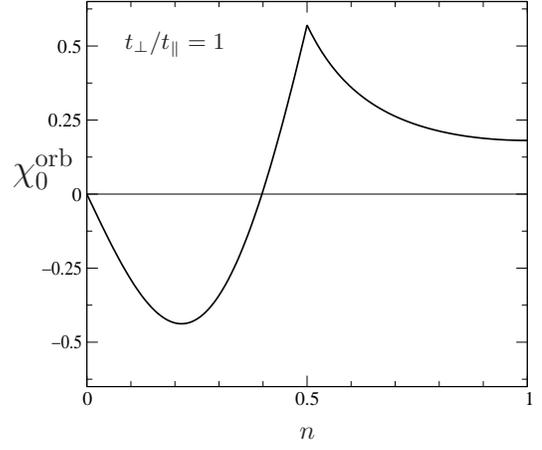}
\caption{Zero field orbital susceptibility of the non-interacting
system for $t_{\perp} = t_{\parallel}$ from
Eqs.~(\ref{eq:susceptibility-1}-\ref{eq:susceptibility-2}). The
singularity at $n = 0.5$ marks the transition from 2 to 4 Fermi
points. When $n < 0.5$, the susceptibility changes sign for $n \simeq
0.3975\ldots$ while the system always has two Fermi points.}
\label{fig:free-susceptibility-n}
\end{figure}

Lastly, the behavior of the zero-field susceptibility
$\chi^{\text{orb}}_0$ can be computed as a function of the electronic
density $n$. With a factor two for the spins, we have:\\ if $\alpha >
1 - \cos \pi n$ (two Fermi points),
\begin{equation}
\label{eq:susceptibility-1} 
\chi_0^{\text{orb}}(n) = -t_{\parallel} \left[ \sin \pi n \left(1 + \frac 1
\alpha \cos \pi n \right ) -\frac \pi \alpha n\right]\,,
\end{equation}
else, if $\alpha < 1 - \cos \pi n$ (four Fermi points), we have
\begin{equation}
\label{eq:susceptibility-2} 
\begin{split}
\chi_0^{\text{orb}}(n) = -t_{\parallel} & \left[
\sqrt{\sin^2\left( \frac{\pi n}{2}\right) - \left( \frac{\alpha}{2}
\right)^2} \left(2 + \frac{\cos \pi n}{\sin^2(\pi n /2)} \right)
\right.\\
&\left. -\frac 2 \alpha \arcsin \left( \frac{\alpha}{2} \frac 1
{\sin(\pi n/2)} \right) \right]\,.
\end{split}
\end{equation}
The curve for $\alpha = 1$ is plotted on
\fig~\ref{fig:free-susceptibility-n} which shows that for $n < 0.5$,
even when there are only two Fermi points, the susceptibility can be
either positive or negative.

\section{Flux quantization and finite size effects}
\label{sec:flux}

In this section, we discuss the quantization of the flux on a finite
size ladder. First, Hamiltonian (\ref{eq:hamiltonian}) clearly gives
$E(\phi) = E(2\pi - \phi)$ so that we can restrict ourselves to the
window $\phi \in [0,\pi]$.  Similarly, from
Eqs.~(\ref{eq:screening-currents}) we have $j(\phi) = -j(2\pi - \phi)$
which implies that $j(0) = j(\pi) = 0$. What is quantization of the
flux $\phi$ on a finite system ? Using periodic boundary conditions
with the gauge of \fig~\ref{fig:schema}, the integrated flux along a
leg is $\pm \phi/2 \times L$ so that there is no remanent flux through
the \emph{cylinder hole} of the periodic ladder if
\begin{equation}\label{eq:flux-quantization}
\phi = m  4\pi/L
\end{equation}
with $m$ an integer. Actually, this can also be simply understood from
momentum quantization $k = 2\pi m/ L$ and looking at the dispersions
$-2t_{\parallel}\cos(k \pm \phi/2)$ on each leg when $t_{\perp} =
0$. This quantization can be checked numerically with exact
diagonalization. Furthermore, another possible gauge which gives the
same flux per plaquette is to take $\phi_{\perp}(x) = \phi x$, with
$\phi_{\perp}$ the flux along a rung, and no flux along the legs. We
can show that the two gauges are strictly equivalent on a finite
system with periodic boundary conditions only if
(\ref{eq:flux-quantization}) is satisfied.

\begin{figure}[t]
\centering
\includegraphics[width=0.9\columnwidth,clip]{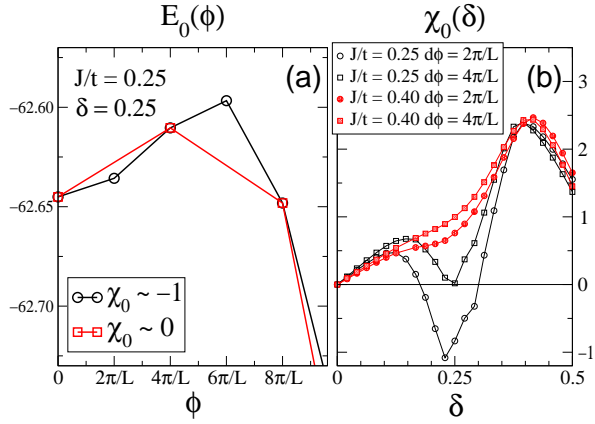}
\caption{(Color online) Finite size effects on the calculation of the
zero-field susceptibility at low $J/t$. {\bf(a)} energies as a
function of flux. {\bf(b)} $\chi_0$ as a function of the doping
$\delta$ from Eq.~(\ref{eq:finite-size-effect}) for two different
$d\phi$.}
\label{fig:finite-size-effects}
\end{figure}

With DMRG, we are using open boundary conditions for which we expect
similar effects due to momentum quantization. For most quantities and
parameters $J/t$, the points with $\phi = 2\pi m /L$ interpolates
nicely with the ones using (\ref{eq:flux-quantization}) so we can
relax the constraint (see \fig~\ref{fig:comparison} for
instance). However, when taking the derivatives such as in
(\ref{eq:screening-currents}) at low $J/t$ where more finite size
effects are present, one must strictly obey
(\ref{eq:flux-quantization}) to have correct estimates (this is
necessary in \fig~\ref{fig:zero-field-susceptibility} for
instance). On \fig~\ref{fig:finite-size-effects}, because we know that
$j_{\parallel}(0) = 0$, the zero field susceptibility is
\begin{equation}
\chi_0 = (E_0(2d\phi) - E_0(0))/(2d\phi^2)\,.
\label{eq:finite-size-effect}
\end{equation}
Taking $d\phi = 2\pi / L$ gives a negative susceptibility while $d\phi
= 4\pi / L$ gives $\chi_0 \sim 0$. Finite size effects are smaller
away from the $\delta = 0.25$ commensurability and also at larger
$J/t$ and .

The DMRG simulations were performed using the standard finite system
algorithm on systems ranging from $L=32$ to $L=64$, with minor
modifications to treat complex wavefunctions. The density matrix at
each step was the sum of the density matrices constructed using the
real and imaginary parts separately. Typically, we kept $m=1400$
states per block, giving a typical truncation error of $10^{-6}$. The
presence of the magnetic field did not increase the truncation error
notably. Correlation functions are computed by averaging two-point
correlations of equal distance. Correlations with one point too close
to one of the edges are removed. Even if there is no translation
symmetry with open boundary conditions, this method gives comparable
results with the one using one point fixed at the middle (but this
method gives access to a larger distance $x$).

\section{Bosonization conventions}
\label{sec:bosonization}

We use the same conventions as in
Ref.~~\onlinecite{Giamarchi2004}. The bosonization procedure starts
from the linearization of the band dispersion in the vicinity the Fermi
points. When there are four Fermi points, two of then in  the up band
and the two other in the down band down
bands (corresponding to the low-field C1S0 phase), we use:
\begin{equation*}
\begin{pmatrix}
    c_1 \\
    c_2 \\
  \end{pmatrix}
= \frac 1 {\sqrt{L}} \sum_k e^{ikx}
 \begin{pmatrix}
    a_k & b_k \\
    b_k & -a_k \\
  \end{pmatrix}
  \begin{pmatrix}
    c_{k,d} \\
    c_{k,u} \\
  \end{pmatrix}\,,
\end{equation*}
with implicit spin index if not explicitly required. We denote by
$\psi_{R/L,d/u}$ the bosonized right and left movers inside  each
bands. Note that we have different Fermi levels $k_{F,d} \equiv k_d
\neq k_{F,u} \equiv k_u$. From (\ref{eq:coherences-factor-a}) and
(\ref{eq:coherences-factor-b}), we deduce that $a_{-k_{d/u}} =
b_{k_{d/u}} \equiv b_{d/u}$ and $b_{-k_{d/u}} = a_{k_{d/u}} \equiv
a_{d/u}$. The bosonized version of the local fermion operators depends
on how many Fermi points we have and which bands are filled. If we
have four Fermi points and overlapping bands, we use
\begin{eqnarray*}
c_{1}(x)/\sqrt{a} \rightarrow&& a_d e^{ik_dx} \psi_{d,R}(x) + b_d e^{-ik_dx} \psi_{d,L}(x)\\
                  &+&  b_u e^{ik_ux} \psi_{u,R}(x) + a_u e^{-ik_ux} \psi_{u,L}(x)\\
c_{2}(x)/\sqrt{a} \rightarrow&& b_d e^{ik_dx} \psi_{d,R}(x) + a_d e^{-ik_dx} \psi_{d,L}(x)\\
                  &-&  a_u e^{ik_ux} \psi_{u,R}(x) - b_u e^{-ik_ux} \psi_{u,L}(x)
\end{eqnarray*}
where the right and left moving Fermi fields have the bosonized representation
\begin{equation*}
\psi_{p,r,\sigma} = \frac{\eta^r_{p, \sigma}}{\sqrt{2 \pi \alpha}}
e^{i \epsilon_r \phi_{r,p, \sigma} }
\end{equation*}
with $\alpha$ a cutoff (not $t_{\perp}/t_{\parallel}$) and $r = R/L$,
$p = d/u$ and $\epsilon_{R/L} = \mp 1$. $\eta^r_{p,\sigma}$ are Klein
factors than ensure anticommutation of  fermion operators
having different spin or band index. The fields $ \phi_{r,p, \sigma}$
are chiral boson fields. 
The non-chiral bosons fields are
defined by: 
\begin{eqnarray}
\phi_{p, \sigma}  &=& [\phi_{L,p,\sigma}+\phi_{R,p,\sigma} ]/2,\\
\theta_{p,\sigma} &=& [\phi_{L,p,\sigma}-\phi_{R,p,\sigma} ]/2,
\end{eqnarray}
and they satisfy commutation relations $[\phi_{p,
  \sigma}(x),\theta_{p',\sigma'}(x')] = i \delta_{pp'}
\delta_{\sigma,\sigma'} \delta(x-x')$.  
As usual in the framework of two coupled chains, we also introduce the
following combinations of the $\phi$ and $\theta$ fields: the charge
and spin modes in each bands $p$ are
\begin{eqnarray}
\phi_{c,p} &=& [\phi_{p,\ups} + \phi_{p,\downs}]/{\sqrt{2}}\\
\phi_{s,p} &=& [\phi_{p,\ups} - \phi_{p,\downs}]/{\sqrt{2}}
\end{eqnarray}
and similar transformations for the $\theta$. And lastly, the $\pm$
combinations
\begin{equation}
\phi_{c/s,\pm} = [\phi_{c/s,d} \pm \phi_{c/s,u}]/{\sqrt{2}}
\end{equation}
The Luttinger parameters associated with these bosons are $K_{c\pm}$
for the charge sectors and $K_{s\pm}$ for the spin sectors.

In the case of two Fermi points (intermediate flux C1S1 phase) and $n
< 1$, only the down band is filled and we can use the results of a
single chain but using
\begin{eqnarray*}
c_{1}(x)/\sqrt{a} &\rightarrow& a_d e^{ik_dx} \psi_{d,R}(x) + b_d e^{-ik_dx} \psi_{d,L}(x)\\
c_{2}(x)/\sqrt{a} &\rightarrow& b_d e^{ik_dx} \psi_{d,R}(x) + a_d e^{-ik_dx} \psi_{d,L}(x)\,.
\end{eqnarray*}
We simply denote by $K_c$ and $K_s$ the Luttinger parameters
corresponding to the charge and spin modes.

In the high-flux C1S0 phase, four points are present in the down
band. With the notation $k_{F,1,d} \equiv k_1 \neq k_{F,2,d} \equiv
k_2$, we have after linearizing the band structure:
\begin{eqnarray*}
c_{1}(x)/\sqrt{a} \rightarrow && a_1 e^{ik_1x} \psi_{1,R}(x) + b_1 e^{-ik_1x} \psi_{1,L}(x)\\
                             &+& a_2 e^{ik_2x} \psi_{2,L}(x) + b_2 e^{-ik_2x} \psi_{2,R}(x)\\
c_{2}(x)/\sqrt{a} \rightarrow && b_1 e^{ik_1x} \psi_{1,R}(x) + a_1 e^{-ik_1x} \psi_{1,L}(x)\\
                             &+& b_2 e^{ik_2x} \psi_{2,L}(x) + a_2 e^{-ik_2x} \psi_{2,R}(x)\,,
\end{eqnarray*}
and similar expressions for the Fermi operators $\psi_{p,r}(x)$.

\section{Diamagnetic susceptibility} 
\label{sec:susceptibility}

Let us consider the Hamiltonian~(\ref{eq:hubbard}) or
(\ref{eq:tJ-hamiltonian}) in the limit $\phi \to 0$. By expanding to
second order, we have:
\begin{equation*}
\Ham = \Ham(\phi=0) - \frac{\phi}{2} L(j_1 - j_2)-\frac{\phi^2}{8} L(K_1 + K_2), 
\end{equation*}
where $j_{1,2}$ are the densities of current operators along chains
$1$ and $2$ and $K_{1,2}$ represent the densities of kinetic energy in
chains $1$ and $2$. Note that these operators are taken at $\phi=0$.
We obtain the density of screening current operator $j_{\parallel} =
-\frac 1 L \partial \Ham / \partial \phi$ as:
\begin{equation*}
j_{\parallel} = \frac 1 2 (j_1-j_2) + \frac{\phi}{4} (K_1+K_2) \,.
\end{equation*}
Using linear response theory, we obtain the expectation value of this
current in this limit as:
\begin{equation*}
j_{\parallel}(\phi) = \moy{j_{\parallel}}_0 = \frac{\phi}{4} \left[ \moy{\moy{ (j_1-j_2);
(j_1-j_2) }}_0 + \moy{K_1+K_2}_0\right] \,, 
\end{equation*}
\noindent where $ \moy{\moy{;}}_0$ represents the retarded response
function and $\moy{}_0$ is the expectation value in the ground state
without magnetic field. In the absence of interchain hopping, the
cross response function $\moy{\moy{j_1;j_2}}_0$ would vanish and
$\moy{j_\parallel}_0$ would be simply the sum of Drude weights of each
chain. The expression of $\moy{j_\parallel}_0$ can be rearranged by
noting that:
\begin{equation}
 j_1 - j_2 = 2\int^x j_\perp 
\end{equation}
as a consequence of Kirchhoff's law. So we have that:
\begin{equation}
\label{eq:general-orbital}
\chi_0 = \int dx \moy{ j_\perp(x) j_\perp(0) }_0 + \frac 1 4 \moy{ K_1+K_2 }_0 \,.
\end{equation}
In the case of negligible transverse current correlations, this term
reduces to the expectation value of the kinetic energy. In an
insulator, this yields  a vanishing diamagnetic susceptibility.

\end{document}